\documentclass{ws-mpla}

\begin{document}

\markboth{Thiago Prud\^encio \& Diego Julio Cirilo-Lombardo \& Humberto Belich}
{Black hole qubit correspondence from quantum circuits}

\catchline{}{}{}{}{}

\title{BLACK HOLE QUBIT CORRESPONDENCE FROM QUANTUM CIRCUITS}

\author{\footnotesize THIAGO PRUD\^ENCIO}

\address{Coordination in Science and Technology - CCCT, Federal University of Maranh\~ao - UFMA,
Campus Bacanga, 65080-805, Sao Lu\'is-MA, Brazil.
thprudencio@gmail.com}

\author{DIEGO JULIO CIRILO-LOMBARDO}

\address{Bogoliubov Laboratory of Theoretical Physics, Joint Institute for Nuclear Research 141980,
Dubna-Moscow, Russia.}

\author{EDILBERTO O. SILVA}

\address{Department of Physics, Federal University of Maranh\~ao - UFMA,
Campus Bacanga, 65085-580 S\~ao Lu\'is-MA, Brazil.}

\author{HUMBERTO BELICH}

\address{Department of Physics and Chemistry, Federal University of Espirito Santo - UFES, 29060-900, Vitoria-ES, Brazil.}

\maketitle

\pub{Received (Day Month Year)}{Revised (Day Month Year)}

\begin{abstract}
We consider the black hole qubit correspondence (BHQC) from quantum
circuits, taking into account the use of gate operations with base in the
formulation of wrapped brane qubits. We interpret these quantum circuits
with base on the BHQC classification of entanglement classes and apply in
specific examples as the generation of Bell, GHZ states, quantum circuit
teleportation and consider the implementation of interchanges in SUSY, black
hole configurations, Freudenthal and rank system constructions. These
results are discussed from the superstring viewpoint showing that the
importance of the construction of the physical states formed by the
entanglement of geometrical entities by cohomological operations
automatically allows the preservation of different amounts of SUSY in the
compactification process given an alternative to the case when fluxes are
introduced in the game: the generalized Calabi-Yau of Hitchin.
\keywords{extra dimensions \and black hole \and qubit correspondence}
\end{abstract}

\ccode{PACS Nos.: 03.67.-a, 02.40.-k, 03.65.Ta}

\section{Introduction}

\label{intro} One of the crucial ingredients to the scenario of extra
dimensions is a brane on which particles of standard model are localized. In
string theory, fields can naturally be localized on D-branes due their open
string endings \cite{polchinski}. Although, extra dimensions are generally
proposed in compactified form, the localization mechanism for gravity can
lead to new possibilities \cite{diego1}. Due to gravitational interactions
between brane in uncompactified 5D-space, a four dimensional Newtonian
behavior can be achieved when the bulk cosmological constant and the brane
tension are related. Additionally, gravitons can be localized on branes by
separating two patches of AdS5 space-time \cite{r}. Static 4D-brane
universes can exist under the requirement that their tension is fine-tuned
with the bulk cosmological constant. Interesting models in which extra
dimensions can also take advantage in the AdS/CFT correspondence, where
strongly coupled 4D theory to 5D warped dimensions can be related \cite{gab}%
. Extensions to charged branes and thermal strings are also possible
scenarios \cite{bel,bel2}. A more recent fact is the string-theoretic
interpretation of the black holes in terms of Dp-branes wrapping around six
compactified dimensions associated to qubits from quantum information (QI) 
\cite{borsten}. This is the so-called black hole qubit correspondence (BHQC) 
\cite{borsten11}. It has lead to important achievements as the association
between black hole entropy emerging from the solution of N = 2 supergravity
STU model of string compactification and tripartite entanglement measurement 
\cite{kallosh,duff}, the association between the black hole configurations
in STU supergravity and entanglement state classification \cite%
{borsten,borsten13} and the identification of the Hilbert space of the
qubits associated to the wrapped branes inside the cohomology of the extra
dimensions \cite{levay}. In fact, many important results were obtained \cite%
{levay4,borsten4,levay3,levay5,levay6,borsten5,borsten2,levay7} (see \cite{borsten12} for a
more complete review). It is believed that this BHQC can
also be extended to the context of supergeometries \cite%
{diego4,diego5,diego6,diego7}.

In this paper, we propose a BHQC from the point of view of quantum circuits
and give the corresponding interpretation. Although recently the role of
entanglement and superpositions were explored in BHQC, quantum circuits was
not explored clearly in this context, mainly the interpretation in the
string side of the BHQC. As we will discuss, this step is fundamental to
explore the BHQC in terms of quantum circuits. We explore the correspondence
and its association to quantum information, giving a clear interpretation
for complete correspondence with BHQC, moving steps forward to the role of
quantum circuits in the BHQC, in particular to explore string theoretical
scenarios. In agreement with previous proposals, we first associate the
wrapped brane qubits in the BHQC according to an one-to-one association \cite{borsten} and then build the necessary 
gate operations to implement quantum circuits \cite{got,chuang}. As instances, we obtain specific quantum
circuits as the generation of Bell states, quantum teleportation circuit 
\cite{benet,brassard} and generation of GHZ states. The fundamental point we
propose the interpretation of these quantum circuits with base on the BHQC
for the classification of entanglement classes \cite{borsten,borsten13}.
Although this clear structure is dressed in a quantum information formalism,
the presence of the interpretation gives the key point for establishing the
BHQC in this context. This step was not clarified in the previous proposals.

We organized the paper as follows: In Sec. II, we consider gate operations
with base on the BHQC of wrapped brane qubits and perform quantum circuits
in BHQC. In Sec. III, we address the BHQC of quantum circuits in the
interpreting these systems from the STU black hole side. Section IV\ is
devoted to make some brief remark concerning the superstring viewpoint of \
the results. Finally, our conclusions are reserved to Sec.V.

\section{BHQC for qubits from extra dimensions, generation of Bell states
and quantum teleportation}

\label{sec:1} The relations in the BHQC for qubits from extra dimensions
start with the associations between one-forms and one-mode states $\Omega
\leftrightarrow |0\rangle, \bar{\Omega} \leftrightarrow |1\rangle$, where
the vacuum state is can be associated to an holomorphic three-form in the
Calabi-Yau space and $K$ is the K\"ahler potential \cite{levay}. The
orthonormality relations are written as 
\begin{eqnarray}
\int_{T^{2}}\Omega \wedge *\bar{\Omega} &\leftrightarrow& \langle
0|0\rangle=1, \\
\int_{T^{2}}\bar{\Omega} \wedge *\Omega &\leftrightarrow& \langle
1|1\rangle=1, \\
\int_{T^{2}}\bar{\Omega} \wedge *\bar{\Omega} &\leftrightarrow& \langle
1|0\rangle= 0, \\
\int_{T^{2}}\Omega \wedge *\Omega &\leftrightarrow& \langle 0|1\rangle = 0.
\end{eqnarray}
The action of the Hodge star operator $*$, that introduces a phase term on $%
|0\rangle$, reads $*|0\rangle = -|0\rangle, *|1\rangle = |1\rangle$. In a
superposed state, $*(|1\rangle \pm |0\rangle)= |1\rangle \mp |0\rangle$. On
the other hand, the action of the flat K\"ahler covariant derivative $D_{%
\hat{\tau}}\Omega=(\bar{z}^{\tau}-z^{\tau}) \left(\partial_{\tau} + \frac{1}{%
2}\partial_{\tau}K \right)\Omega$ follow the rules $D_{\hat{\tau}}\Omega = 
\bar{\Omega}$, $D_{\hat{\tau}}\bar{\Omega} = 0$, and the flat adjoint
covariant derivatives follows the corresponcences $D_{\hat{\bar{\tau}}%
}\Omega = 0$ and $D_{\hat{\bar{\tau}}}\bar{\Omega}=\Omega$, leading to BHQC
with bit-flippers $D_{\hat{\tau}}\Omega \leftrightarrow \uparrow|0\rangle =
|1\rangle$, $D_{\hat{\tau}}\bar{\Omega} \leftrightarrow \uparrow|1\rangle =
0 $, $D_{\hat{\bar{\tau}}}\Omega \leftrightarrow \downarrow|0\rangle = 0$, $%
D_{\hat{\bar{\tau}}}\bar{\Omega} \leftrightarrow
\downarrow|1\rangle=|0\rangle.$ A general qubit state in extra dimensions
can then be represented by a non-normalized qubit $|\Gamma\rangle=
\alpha|1\rangle + \beta |0\rangle$. The basic elements $*$, $\uparrow$, $%
\downarrow$, $|1\rangle$, $|0\rangle$ can be used to implement the BHQC to a
large range of combinations. The Hodge star operators and the covariant
derivatives can be combined to define operators 
\begin{eqnarray}
\lambda_{1} &=& *\uparrow + \uparrow *, \\
\lambda_{2} &=& *\downarrow + \downarrow *, \\
\lambda_{3} &=& *\downarrow + \uparrow *, \\
\lambda_{4} &=& *\uparrow + \downarrow *.
\end{eqnarray}
These operators act on one-mode states $|0\rangle$ and $|1\rangle$ leading
to the following relations $\lambda_{1}|j\rangle = 0, \lambda_{2}|j\rangle =
0, \lambda_{3}|j\rangle = -|j\oplus 1\rangle, \lambda_{4}|j\rangle =
|j\oplus 1\rangle, $
where $j=0,1 \in Z_{2}$. 
An immediate consequence of this operation is $\lambda_{3}^{2}%
\leftrightarrow I, \lambda_{4}^{2}\leftrightarrow I$. 
The action of these operators on qubit states are $\lambda_{3}\left(\alpha|0%
\rangle + \beta|1\rangle\right)= -\left(\alpha|1\rangle +
\beta|0\rangle\right)$ and $\lambda_{4}\left(\alpha|0\rangle +
\beta|1\rangle\right)= \left(\alpha|1\rangle + \beta|0\rangle\right)$. These
operators are then equivalent to a NOT gate \cite{chuang} and are related to
each other by means of $\lambda_{4}=-\lambda_{3}$. 
It follows, all the one-mode gate operations can be realized by combinations
of the actions of the operator $\lambda_{4}$ and its square $%
\lambda_{4}^{2}=I$. We can also have Hadamard gates by means of the
operations %
$\left(I + *\lambda_{4} \right)|1\rangle = |1\rangle -|0\rangle$ and $%
\left(I + *\lambda_{4} \right)|0\rangle = |0\rangle + |1\rangle$ or 
$\lambda_{4}\left(I + *\lambda_{4} \right)|1\rangle = |0\rangle -|1\rangle$
and $\lambda_{4}\left(I + *\lambda_{4} \right)|0\rangle = |0\rangle +
|1\rangle$. 
A $\sigma_{2}$-type gate can be obtained from 
$\lambda_{4} * |0\rangle = -|1\rangle$ and $\lambda_{4} * |1\rangle =
|0\rangle$ or $\lambda_{3} * |0\rangle = |1\rangle$ and $\lambda_{3} *
|1\rangle = -|0\rangle$. In similar fashion, other gate operations can be
built from suitable applications, leading to a BHQC for a universal quantum
computation.

The correspondence can be generalized to $n$-mode case. In the case of a
two-mode space, whose basis is $\lbrace |00\rangle, |10\rangle, |01\rangle,
|11\rangle \rbrace$, two mode gate operation can be obtained from one-mode
ones. For instance, 
\begin{eqnarray}
* \otimes * |ij\rangle = \left\{ 
\begin{array}{ll}
|ii\rangle, & \mbox{ if } i=j; \\ 
-|ij\rangle, & \mbox{ otherwise.}%
\end{array}
\right.
\end{eqnarray}
\begin{eqnarray}
\uparrow \otimes \uparrow |ij\rangle = \left\{ 
\begin{array}{ll}
|11\rangle, & \mbox{ if } i=j= 0; \\ 
0, & \mbox{ otherwise.}%
\end{array}
\right.
\end{eqnarray}
\begin{eqnarray}
\downarrow \otimes \downarrow |ij\rangle = \left\{ 
\begin{array}{ll}
|00\rangle, & \mbox{ if } i=j= 1; \\ 
0, & \mbox{ otherwise.}%
\end{array}
\right.
\end{eqnarray}
It is also easy to check that $\uparrow \otimes \downarrow|01\rangle =
|10\rangle$ and $\downarrow \otimes \uparrow |10\rangle = |01\rangle$, zero
otherwise. 
As in the one-mode case, we can define new operators 
\begin{eqnarray}
\Lambda_{1} &=& *\otimes\uparrow + \uparrow \otimes *, \\
\Lambda_{2} &=& *\otimes\downarrow + \downarrow \otimes *, \\
\Lambda_{3} &=& *\otimes\uparrow + \downarrow \otimes *, \\
\Lambda_{4} &=& *\otimes\downarrow + \uparrow \otimes *,
\end{eqnarray}
that lead to the following results $\Lambda_{1}|00\rangle =
-\left(|01\rangle + |10\rangle\right)$, $\Lambda_{1}|11\rangle = 0$, $%
\Lambda_{1}|01\rangle = |11\rangle$, $\Lambda_{1}|10\rangle = |11\rangle$, $%
\Lambda_{2}|00\rangle = 0$, $\Lambda_{2}|11\rangle=|01\rangle + |10\rangle$, 
$\Lambda_{2}|01\rangle = -|00\rangle$, $\Lambda_{2}|10\rangle = -|00\rangle$%
, $\Lambda_{3}|00\rangle = -|01\rangle$, $\Lambda_{3}|11\rangle = |01\rangle$%
, $\Lambda_{3}|01\rangle = 0$, $\Lambda_{3}|10\rangle = |11\rangle
-|00\rangle$, $\Lambda_{4}|00\rangle = -|01\rangle$, $\Lambda_{4}|11\rangle
= |01\rangle$, $\Lambda_{4}|01\rangle = 0$, $\Lambda_{4}|10\rangle =
|11\rangle -|00\rangle$. In particular, we can verify $\Lambda_{2}%
\Lambda_{1}|00\rangle=2|00\rangle$, $\Lambda_{1}\Lambda_{2}|11\rangle=2|11%
\rangle$.

A controlled not (CNOT) gate can be implemented considering the conjugation
rules 
\begin{eqnarray}
\Omega\otimes\Omega &\rightarrow& \Omega\otimes\Omega,  \label{om1} \\
\Omega\otimes\bar{\Omega} &\rightarrow& \Omega\otimes\bar{\Omega},
\label{om2} \\
\bar{\Omega}\otimes\Omega &\rightarrow& \bar{\Omega}\otimes\bar{\Omega},
\label{om3} \\
\bar{\Omega}\otimes\bar{\Omega} &\rightarrow& \bar{\Omega}\otimes\Omega,
\label{om4}
\end{eqnarray}
These rules corresponds to the operation 
\begin{eqnarray}
|ij\rangle \rightarrow |i\rangle|i\oplus j\rangle,
\end{eqnarray}
and then implement a CNOT gate, where $i$ corresponds to the control and $j$
to the target. As can be verified, this gate can be written as the action of
the following operator $U_{CNOT}(i)=(I\otimes \lambda_{4})^{i}$. 

As direct application of the BHQC, we can verify the usual quantum circuits
implementing Bell states and quantum teleportation. Taking the input state $%
|00\rangle$, we can apply a set of gate operations in order to generate all
the Bell states. We apply the gate operation $(*\otimes *)\Lambda_{1}$ on
the input state we have the first Bell state 
\begin{eqnarray}
|B_{1}\rangle=(*\otimes *)\Lambda_{1}|00\rangle &=& |01\rangle + |10\rangle,
\end{eqnarray}
The second Bell state $|B_{2}\rangle$ can be generated by applying the $%
(I\otimes *)$ gate operation 
\begin{eqnarray}
|B_{2}\rangle=(I\otimes *)|B_{1}\rangle &=& |01\rangle - |10\rangle.
\end{eqnarray}
The application of $(I\otimes \uparrow)$ leads to 
\begin{eqnarray}
|B_{3}\rangle=(I\otimes \uparrow)|B_{2}\rangle &=& |00\rangle - |11\rangle.
\end{eqnarray}
Next, applying $(I\otimes *)$, results 
\begin{eqnarray}
|B_{4}\rangle=(I\otimes \lambda_{4}\lambda_{3})|B_{3}\rangle &=& |00\rangle
+ |11\rangle.
\end{eqnarray}
\begin{figure}[h]
\centering
\includegraphics[scale=0.35]{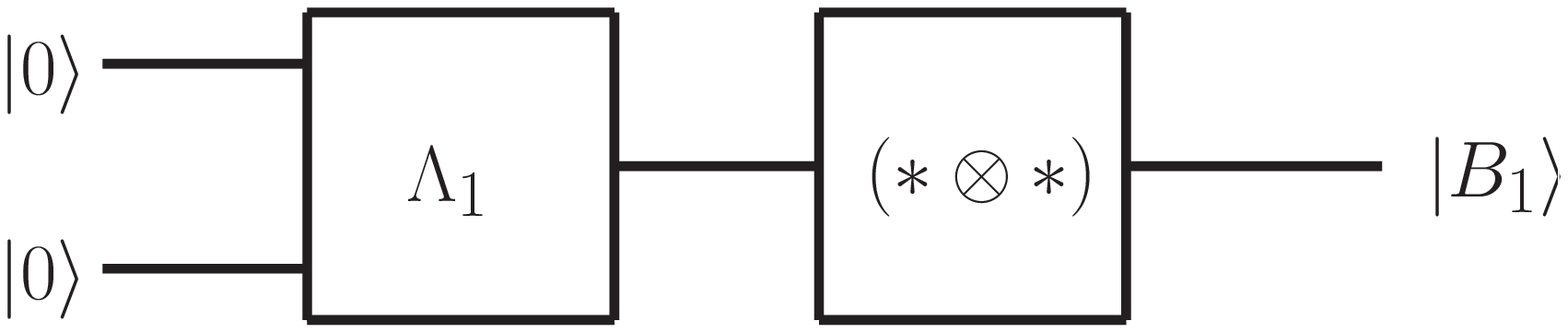} \includegraphics[scale=0.35]{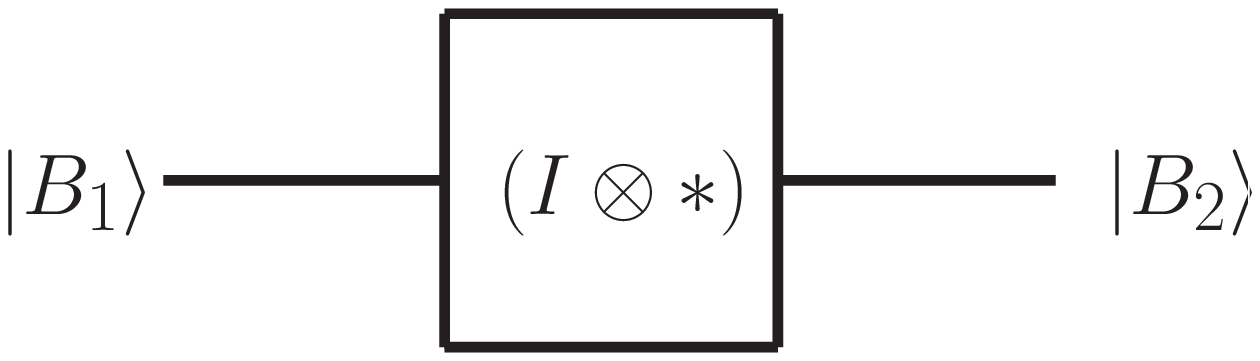} %
\includegraphics[scale=0.35]{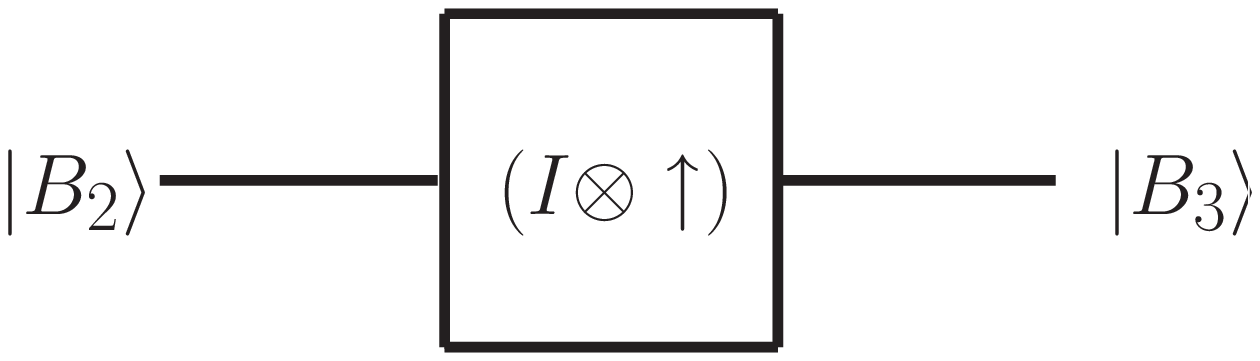} \includegraphics[scale=0.35]{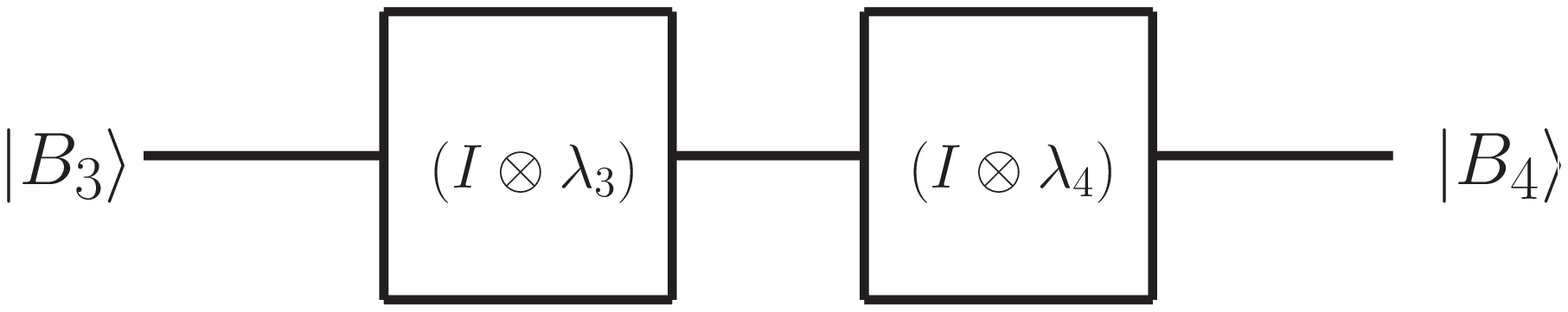}
\caption{(Color online) Quantum circuit for the generation of the bell
states from qubeds.}
\label{b1}
\end{figure}

Let us consider now a qubit state with coefficients $\alpha$ and $\beta$
unknown. We can realize a quantum state teleportation by means of a circuit
teleportation \cite{brassard} with the use of the previous gate operations.
The initial state can be represented by 
\begin{eqnarray}
|\Gamma\rangle_{a}|B_{4}\rangle_{b_{1}b_{2}},  \label{2wwe}
\end{eqnarray}
where $b_{1}$ correspond to the first mode and $b_{2}$ to the second mode of
the Bell state. The coefficients of $|\Gamma\rangle_{a}=\alpha|0\rangle_{a}+%
\beta|1\rangle_{b}$ are generally unknow.

\begin{figure}[h]
\centering
\includegraphics[scale=0.32]{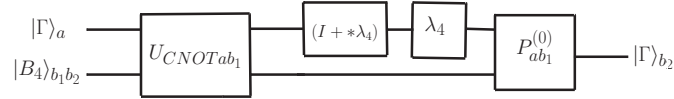}
\caption{(Color online) Sequence of quantum gate operations in the quantum
circuit teleportation for qubeds.}
\label{t1}
\end{figure}

Under a CNOT gate where the qubed mode $a$ is the control state, the state (%
\ref{2wwe}) is modified to 
\begin{eqnarray}
&&\left( \alpha |00\rangle _{ab_{1}}+\beta |11\rangle _{ab_{1}}\right)
|0\rangle _{b_{2}}+\left( \alpha |01\rangle _{ab_{1}}+\beta |10\rangle
_{ab_{1}}\right) |1\rangle _{b_{2}}.  \nonumber \\
&&
\end{eqnarray}%
Applying the Hadamard gate operation in the mode $a$, we arrive at 
\begin{eqnarray}
&&\left[ \alpha \left( |0\rangle _{a}+|1\rangle _{a}\right) |0\rangle
_{b_{1}}+\beta \left( |1\rangle _{a}-|0\rangle _{a}\right) |1\rangle _{b_{1}}%
\right] |0\rangle _{b_{2}}  \nonumber \\
&+&\left[ \alpha \left( |0\rangle _{a}+|1\rangle _{a}\right) |1\rangle
_{b_{1}}+\beta \left( |1\rangle _{a}-|0\rangle _{a}\right) |0\rangle _{b_{1}}%
\right] |1\rangle _{b_{2}},  \nonumber \\
&&
\end{eqnarray}%
and applying a $\lambda _{4}$ operation on the mode $a$ we arrive in the
opposed Hadamard operation 
\begin{eqnarray}
&&\left[ \alpha \left( |0\rangle _{a}+|1\rangle _{a}\right) |0\rangle
_{b_{1}}+\beta \left( |0\rangle _{a}-|1\rangle _{a}\right) |1\rangle _{b_{1}}%
\right] |0\rangle _{b_{2}}  \nonumber  \label{diao} \\
&+&\left[ \alpha \left( |0\rangle _{a}+|1\rangle _{a}\right) |1\rangle
_{b_{1}}+\beta \left( |0\rangle _{a}-|1\rangle _{a}\right) |0\rangle _{b_{1}}%
\right] |1\rangle _{b_{2}}.  \nonumber \\
&&
\end{eqnarray}%
Since the usual projection relations applies to the one-forms, we can
project the total state into the state $|00\rangle _{ab_{1}}$, given by the
application of the projector 
\[
P_{ab_{1}}^{(0)}=\left( |00\rangle \langle 00|\right) _{ab_{1}},
\]%
on the whole state (\ref{diao}), the mode $b_{2}$ then assumes the state 
\[
\alpha |0\rangle _{b_{2}}+\beta |1\rangle _{b_{2}},
\]%
which corresponds to the quantum state teleportation of the qubed from the
mode $a$ to $b_{2}$ (figure \ref{t1}).

\section{Quantum circuits: BHQC interpretation}

\begin{figure}[h]
\centering
\includegraphics[scale=0.4]{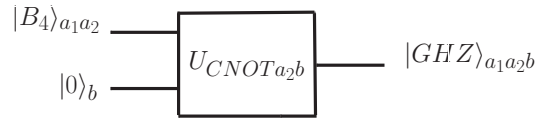}
\caption{(Color online) Quantum circuit generating a GHZ state.}
\label{tghz}
\end{figure}

The association between the entropy of an STU black hole supergravity and a
3-tangle of a given tripartite state in one-to-one correspondence \cite%
{borsten} leads to a clear association with a purelly tripartite entangled
state, the GHZ state \cite{kallosh}. Considering a three qubit quantum
circuit, this state is generated from a Bell state $|B_{4}%
\rangle_{a_{1}a_{2}}$ and a third state $|0\rangle_{b}$, applying a CNOT
gate in $a_{2}b$, where $a_{2}$ is the control, we generate a GHZ state
(figure \ref{tghz}). 
Alternatively, the same result is obtained if the control qubit is $a_{1}$.
We can just write, $k=1,2$, 
\begin{eqnarray}
U_{CNOT_{a_{k}b}}|B_{4}\rangle_{a_{1}a_{2}}\otimes|0\rangle_{b}&=&
|000\rangle_{a_{1}a_{2}b} + |111\rangle_{a_{1}a_{2}b}.  \nonumber
\end{eqnarray}
This state corresponds to four D3-branes intersecting over a string in the
BHQC, according to the classification of three-qubit states as
supersymmetric black holes \cite{borsten}. In this classification, a BHQC of
quantum circuit can be used to generate different wrapping configurations of
these intersecting D3-branes by the action of appropriate operators. In
fact, if we consider a more general state corresponding to a STU black hole $%
|ABC\rangle =\sum_{ijk=0}^{1}\alpha_{ijk}|ijk\rangle$, in particular, the
triality interchanges and class changes can be implemented using BHQC of
quantum circuits. In particular, the classes $A-B-C$, $A-BC$, $W$ and $GHZ$
as described in \cite{borsten} can be interconnected by gate operations. We
can implement for instance the an interchange of $A-B-C$ to $A-BC$ by
applying a Hadamard gate in the third qubit and then a CNOT gate operations
between the second and the third as control 
\begin{eqnarray}
|000\rangle \rightarrow |000\rangle + |001\rangle \rightarrow |101\rangle +
|110\rangle.  \nonumber
\end{eqnarray}

\begin{figure}[h]
\centering
\includegraphics[scale=0.4]{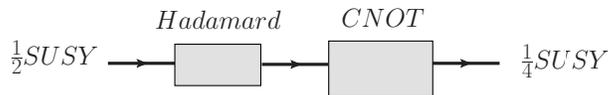}
\caption{(Color online) Change in the SUSY configuration of a small black
hole corresponding to a BHQC quantum circuit.}
\label{tsusy}
\end{figure}

This change corresponds to modify the SUSY configuration of the small black
hole from $1/2$ preserved to $1/4$ preserved (figure \ref{tsusy}). On the
other hand, the quantum circuit for the generation of a GHZ state
corresponds to a passage from a small (non-attractor) to a large (attractor)
black hole with SUSY $1/8$ preserved or completelly broken (figure \ref%
{tsusy2}). 
\begin{figure}[h]
\centering
\includegraphics[scale=0.4]{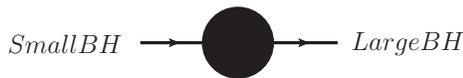}
\caption{(Color online) Change from a small to a large black hole from a
BHQC quantum circuit.}
\label{tsusy2}
\end{figure}

An implementation of STU black holes associated to four qubit systems \cite%
{borsten13,levay4} is made in analogous way to the previous circuits, we can
derive a quantum circuit for four entanglement to implement a BHQC applying
adequate quantum gate operations in the presence of an auxiliary qubit, for
example, $|GHZ\rangle \otimes |0\rangle $ under a CNOT gate. The extension
to quantum circuits with a higher number of qubits can be made
straightforwardly by the introduction of new auxiliary qubits. This
procedure corresponds to deal with a given $n$-form in the cohomology class
of wrapped D$p$-branes. Other mechanisms, as the process of moduli
stabilization at the horizon associated to entanglement distillation to a
GHZ state \cite{levay6} can also be implemented in a more clear form from
the perspective of BHQC in quantum circuits. Because the automorphism group
corresponds to the U-duality group of a variety 4-dimensional
supergravities, in the context of the Freudenthal's construction \cite{vrana}%
, the BHQC can be implemented to start from a three-qubit separable
projective coset and implement the gate operations to move to a biseparable
projective coset or two entangled qubits coset. For instance, {\small 
\begin{eqnarray}
SL(2,C)\times SL(2,C)\times SL(2,C) &\rightarrow &SL(2,C)\times SL(4,C) 
\nonumber \\
&\rightarrow &SL(6,C).  \nonumber
\end{eqnarray}%
} As a consequence, quantum circuits can also be implemented to connect
automorphism (SLOCC) groups in BHQC. In a FTS rank system, it corresponds to
move from given rank system and other rank system, as, for example, starting
from a rank 1 system and generate a rank 4, rank 3 system or rank 2a, 2b, 2c
system. It is important to remind in these cases that, the conventional concept
of matrix rank may be generalised to Freudenthal triple systems \cite{ferrar,krutelevich} in a natural
and authomorphic invariant manner.

\section{Superstring viewpoint}

As is well known Calabi-Yau manifolds are important in superstring theory
because the ten conjectural dimensions are supposed to come as four of which
we are aware, carrying some kind of fibration with fiber dimension six, and
they leave some of the original supersymmetry (SUSY) unbroken \cite{green,ooguri,park}. The importance of the states contructed via
entanglement in the different quantum-information theoretical processes
described here, is that as the starting point, the geometry of a Calabi-Yau
manifold \ is used to define such states. Consequently, the topology of the
Calabi-Yau manifold changes (not the dimension) making that the preserved
SUSY under the compactification process also change. That means that we have
a mechanism to control the preserved susy under the compactification process.

In resume: the importance of the construction of the physical states formed
by the entanglement of geometrical entities by cohomological operations
automatically allows the preservation of different amounts of SUSY in the
compactification process given an alternative to the case when fluxes are
introduced in the game: the generalized Calabi-Yau of Hitchin \cite{hitchin,huybrechts}.

\section{Conclusions}

We have considered quantum circuits implemented in the context of BHQC with
qubits from wrapped branes. Applying first to obtain quantum circuits to
generation of Bell, GHZ states and teleportation, we then connect the states
involved interpreting these quantum circuits in terms of the entanglement
classes classification associated to the entropy and SUSY configurations of
STU black holes \cite{borsten}. As a consequence, we used our formulation to
consider the interchange of SUSY, black hole configurations, Freudenthal's
and rank system contructions by means of BHQC quantum circuits.

As we have clearly seen, the results show that the cohomological operations
performed and proposed in this paper allow the amount of supersymmetry
preserved, be more flexible under compactification that from the quantum
field theoretical point of view is extremely important due that the
interplay between the SUSY\ preserved, the moduly space and the
geometry/topology of the remanent (super) phase space

This proposal is also useful in the mechanism of moduli stabilization at the
horizon \cite{levay6,taylor}, by considering corresponding BHQC quantum
circuits, what can be implemented in a future work elsewhere.

In the context of Riemannian superspaces of \cite{diego6,diego7}, it is
possible to reformulate consistently this construction at the operator level
avoiding the black hole interpretation due that, as we have been shown, the
black/hole entropy argument is not necessary because the spacetime carry
itself the quantum/statistical properties (there is not dependence of such
properties on a particular solution, as in the black hole case). This is
also important in the construction of the physical states formed by the
entanglement of geometrical entities by cohomological operations.

\section{Acknowledgements}

 D.J.C-L. thanks JINR-BLTP
(Russia) for hospitality and financial support. T. P. and E. O. S. acknowledge FAPEMA (Brazil) 
for finantial support. T.P. also thanks Enxoval Project UFMA/PPPG No. 03/2014.

\end{document}